\documentclass[a4paper, 12pt, oneside]{article}

 \usepackage[cp1251]{inputenc}
 \usepackage[english, russian]{babel}
\usepackage{amsmath, latexsym, amssymb, bm, array, graphics, eucal, amsfonts}

\usepackage[dvips]{graphicx}
 \makeatletter

\renewcommand{\Re}{\mathop{\rm Re}}
\renewcommand{\Im}{\mathop{\rm Im\,}}

\makeatother

\begin{document}
\thispagestyle{empty}
\large
\renewcommand{\abstractname}{Abstract}
\renewcommand{\refname}{\begin{center}
 REFERENCES\end{center}}
\newcommand{\mc}[1]{\mathcal{#1}}
\newcommand{\E}{\mc{E}}

\begin{center}
\bf Generation of longitudinal
electric current by transversal electromagnetic field in
Maxwellian plasmas
\end{center} \medskip

\begin{center}
  \bf A. V. Latyshev\footnote{$avlatyshev@mail.ru$} and
  A. A. Yushkanov\footnote{$yushkanov@inbox.ru$}
\end{center}\medskip

\begin{center}
{\it Faculty of Physics and Mathematics,\\ Moscow State Regional
University, 105005,\\ Moscow, Radio str., 10A}
\end{center}\medskip

\begin{abstract}
The analysis of nonlinear interaction of transversal
electromagnetic field with Maxwellian collisionless
classical and quntum plasmas is carried out.
Formulas for calculation electric current in Maxwellian collisionless
classical and quntum plasmas are deduced.
It has appeared, that the nonlinearity account leads to occurrence
of the longitudinal electric current directed along a wave vector.
This second current is orthogonal to the known transversal  current,
received at the classical linear analysis.
Graphic comparison of density of  electric current for classical
Maxwellian plasmas and Fermi---Dirac plas\-mas (plasmas with any
degree of degeneration of electronic gas) is carried out.
Graphic comparison of density of electric current for
classical and quantum Maxwellian plasmas is carried out.
Also comparison of dependence of density of electric
current of quantum Maxwellian plasmas
from dimensionless wave number at various values
of dimensionless frequency of oscillations of
electro\-mag\-netic field is carried out.


electron gas
\end{abstract}

\begin{center}
\bf  Introduction
\end{center}

The nonlinear phenomena in plasma are studied during the long
time \cite{Gins} -- \cite{Lat1}. In work \cite{Mermin} was
the approach to studying of dielectric function in quantum plasma is offered.
Then in our works \cite{Lat2}--\cite{Lat5} was
dielectric permeability in the quantum
collisional to plasma is investigated.

In works \cite{Lat6} and \cite{Lat7} plasma with
any degeneration of electronic gas (plasma of
Fermi---Dirac) was considered . It has been shown, that with use
square-law decomposition of function of distribution is possible to reveal
longitudinal electric current. This current is generated by the
transversal electromagnetic field. Longitudinal current
is perpendicular to the transversal current which is at the linear analysis.

Let us notice, that existence of the longitudinal current generated
by transversal electromagnetic field, it has been noticed more half  century
ago in work \cite{Zyt2}.

In the present work formulas for calculation electric current
into Maxwellian collisionless plasma  at any temperature (at any
degrees of degeneration of the electronic gas)  are deduced.

It has appeared, that electric current expression consists of two sum\-mands.
The first summand, linear on vector potential, is
known classical expression of electric current.
This electric current is directed along vector potential of electromagnetic
field. The second summand represents itself electric current,
which is proportional to the square
vector potential of electromagnetic field. The second current
is perpendicular to the first and it is directed along the  wave
vector.
Occurrence of the second current comes to light the spent account
nonlinear character
interactions of electromagnetic field with plasma.

Then the kinetic equation with Wigner integral
concerning quantum function of distribution is used.
The nonlinear analysis is made and the formula for calculation
longitudinal electric current is deduced. This current also is generated
by transversal electromagnetic field.

\begin{center}
  {\bf 1. Vlasov kinetic equation and its solution}
\end{center}

Let us consider Vlasov equation describing behaviour
of collisionless plasmas
$$
\dfrac{\partial f}{\partial t}+\mathbf{v}\dfrac{\partial f}{\partial
\mathbf{r}}+
e\bigg(\mathbf{E}+
\dfrac{1}{c}[\mathbf{v},\mathbf{H}]\bigg)
\dfrac{\partial f}{\partial\mathbf{p}}=0.
\eqno{(1.1)}
$$

Vector potential we take  as orthogonal to direction of a wave vector
$\mathbf{k}$
$$
\mathbf{k}\mathbf{A}(\mathbf{r},t)=0.
\eqno{(1.2)}
$$
in the form of the running harmonious wave
$$
\mathbf{A}(\mathbf{r},t)=\mathbf{A}_0
e^{i(\mathbf{k} \mathbf{r}-\omega t)}.
$$

Scalar potential we will consider equal to zero.
Electric and magnetic fields are connected with vector potential
by equalities
$$
\mathbf{E}=-\dfrac{1}{c}\dfrac{\partial \mathbf{A}}{\partial t},\qquad
\mathbf{H}={\rm rot} \mathbf{A}.
\eqno{(1.3)}
$$

The wave vector we direct along axis $x$: $\mathbf{k}=k(1,0,0)$,
and vector potential of elec\-tromagnetic field we direct along axis
$y$
$$
\mathbf{A}=A_y(x,t)(0,1,0),\qquad A_y(x,t)\sim e^{i(kx-\omega
t)}.
$$

Then
$$
A_y=-\dfrac{ic}{\omega}E_y,\qquad \mathbf{H}=\dfrac{ck}{\omega}
E_y(0,0,1),
$$
$$
 [\mathbf{v,H}]=\dfrac{ck}{\omega}E_y (v_y,-v_x,0).
$$

Let us operate with  method  of consecutive approximations.
Con\-si\-de\-ring, that the member with an electromagnetic
field has an order, on
unit smaller other members,
let us rewrite the equation (1.1) in the form
$$
\dfrac{\partial f^{(k)}}{\partial t}+v_x\dfrac{\partial f^{(k)}}{\partial x}+
$$
$$
+eE_y\Bigg(\dfrac{\partial f^{(k-1)}}{\partial p_y}
\Big(1-\dfrac{kv_x}{\omega}\Big)+\dfrac{kv_y}{\omega}
\dfrac{\partial f^{(k-1)}}{\partial p_x}\Bigg)=0,\quad k=1,2.
\eqno{(1.4)}
$$

Here in zero approximation $f^{(0)}$ is the absolute
Maxwell---Boltzmann  dist\-ri\-bu\-tion,

$$
f^{(0)}=f_0=
N\Big(\dfrac{\beta}{\pi}\Big)^{3/2}e^{-\beta v^2},\qquad \beta=\dfrac{m}
{2k_BT},
$$
$k_B$ is the Boltzmann constant, $T$ is the plasmas temperature.

It is easy to see, that
$\mathbf{P}=\dfrac{\mathbf{v}}{v_T}=\dfrac{\mathbf{p}}{p_T}$
is the dimensionless  electron velocity (or momentum),
$v_T$ is the thermal velocity of electrons,
$$
v_T=\sqrt{\dfrac{2k_BT}{m}}.
$$

We notice that
$$
[\mathbf{v,H}]\dfrac{\partial f^{(0)}}{\partial \mathbf{p}}=0,
$$
because
$$
\dfrac{\partial f^{(0)}}{\partial \mathbf{p}}\sim \mathbf{v}.
$$

Therefore in first approximation Vlasov equation have the form
$$
\dfrac{\partial f^{(1)}}{\partial t}+v_x\dfrac{\partial f^{(1)}}{\partial x}=
-
eE_y\dfrac{\partial f^{(0)}}{\partial p_y}.
\eqno{(1.5)}
$$

And in second approximation Vlasov equation have the
following  form
$$
\dfrac{\partial f^{(2)}}{\partial t}+v_x\dfrac{\partial f^{(2)}}{\partial x}=
$$
$$
=-
eE_y\Bigg(\dfrac{\partial f^{(1)}}{\partial p_y}\cdot
\dfrac{\omega-kv_x}{\omega}+\dfrac{kv_y}{\omega}
\dfrac{\partial f^{(1)}}{\partial p_x}\Bigg).
\eqno{(1.6)}
$$

We search solution in first approximation in the form
$$
f^{(1)}=f^{(0)}+f_1(x,t,P_x),\qquad f_1\sim E_y(x,t).
$$

In this approximation the equation (1.5) becomes simpler
$$
\dfrac{\partial f_1}{\partial t}+v_TP_x\dfrac{\partial f_1}{\partial x}=
-\dfrac{eE_y}{mv_T}\dfrac{\partial f_0}{\partial P_y}.
$$

From this equation we find
$$
-i(\omega-kv_TP_x)f_1=\dfrac{2eE_y}{p_T}f_0(P)P_y,
$$
from which
$$
f_1=\dfrac{2ieE_y}{p_T}\dfrac{P_y f_0(P)}{\omega-kv_TP_x}.
\eqno{(1.7)}
$$

In the second approximation for function $f^{(2)}$ we search in the
form
$$
f^{(2)}=f^{(1)}+f_2(x,t,v_x),\quad\quad f_2\sim E_y^2(x,t).
$$

Let us substitute $f^{(2)}$ in (1.6). We receive the equation
$$
\dfrac{\partial f_2}{\partial t}+v_x\dfrac{\partial f_2}{\partial x}=-
\dfrac{eE_y}{\omega}\Bigg[(\omega-kv_x)\dfrac{\partial f_1}{\partial p_y}
+kv_y\dfrac{\partial f_1}{\partial p_x}\Bigg].
$$

We transform this equation to dimensionless parameters
$$
\dfrac{\partial f_2}{\partial t}+v_TP_x\dfrac{\partial f_2}{\partial x}=-
\dfrac{eE_y}{\omega p_T}
\Bigg[(\omega-kv_TP_x)\dfrac{\partial f_1}{\partial P_y}
+kv_TP_y\dfrac{\partial f_1}{\partial P_x}\Bigg].
$$

From this equation we find
$$
f_2=\dfrac{e^2E_y^2}{\omega p_T^2(\omega-kv_TP_x)}
\Bigg[\dfrac{ \partial (P_yf_0(P))}{\partial P_y}
+kv_T P_y^2\dfrac{\partial}{\partial P_x}
\Big(\dfrac{f_0(P)}{\omega-kv_TP_x}\Big)\Bigg].
\eqno{(1.8)}
$$

We introduce new parameters $q$ and $\Omega$,  $q$ is the
dimensionless wave number, $\Omega$ is the dimensionless
frequency of elecromagnetic field,
$$
q=\dfrac{k}{k_T}, \qquad \Omega=\dfrac{\omega}{k_Tv_T},\qquad
k_T=\dfrac{p_T}{\hbar}=\dfrac{mv_T}{\hbar}.
$$

Then we rewrite the equation (1.8) in the form
$$
f_2=\dfrac{e^2E_y^2}{p_T^2k_T^2v_T^2(\Omega-qP_x)}
\Bigg[\dfrac{ \partial (P_yf_0(P))}{\partial P_y}
+qP_y^2\dfrac{\partial}{\partial P_x}
\Big(\dfrac{f_0(P)}{\Omega-qP_x}\Big)\Bigg].
\eqno{(1.8')}
$$

Distribution function in square-law approximation on the field
it is constructed
$$
f=f^{(2)}=f^{(0)}+f_1+f_2,
\eqno{(1.9)}
$$
where $f_1, f_2$ are given accordingly by formulas (1.7) and (1.8).

\begin{center}
  \bf 2. Density of electric current
\end{center}

Let us calculate current density
$$
\mathbf{j}=e\int \mathbf{v}f d^3v=ev_T^4\int \mathbf{P}f d^3P=
ev_T^4\int \mathbf{P}f_1 d^3P.
\eqno{(2.1)}
$$

By means of (1.8) it is visible, that the vector of current
density  has two non-zero components
$$
\mathbf{j}=(j_x,j_y,0).
$$

Here $j_y$ is the density of known transversal current, calculated as
$$
j_y=ev_T^4\int P_yf d^3P=ev_T^4\int P_yf_1 d^3P.
\eqno{(2.2)}
$$

This current is directed along electric field, its density
is deduced by means of linear approximation of distribution function.

Square-law on quantity of an electromagnetic field composed $f_2$
the contribution to density of a current does not bring.
Density of transversal current  is calculated under the formula
$$
j_y=\dfrac{2ie^2v_T^2}{mk_T}E_y(x,t)\int\dfrac{P_y^2f_0(P)d^3P}
{\Omega-qP_x}.
\eqno{(2.3)}
$$

Here $k_T$ is the thermal wave number,
$$
k_T=\dfrac{p_T}{\hbar}=\dfrac{mv_T}{\hbar}.
$$

Let us calculate the longitudinal current. For density of
longitudinal current according to definition it is had
$$
j_x=e\int v_xfd^3v=ev_T^4\int P_xf_2d^3P.
$$

Having taken advantage $(1.8')$, from here we receive, that
$$
j_x=\dfrac{e^3E_y^2}{k_T^2m^2\Omega}\int
\Bigg[\dfrac{\partial(P_yf_0(P))}{\partial P_y}
+qP_y^2\dfrac{\partial}{\partial P_x}
\Big(\dfrac{f_0(P)}{\Omega-qP_x}\Big)\Bigg]\dfrac{P_xd^3P}{\Omega-qP_x}.
\eqno{(2.4)}
$$

The first integral from (2.4) is equal to zero. Really, we will consider
internal integral on $P_y $
$$
\int\limits_{-\infty}^{\infty}
\dfrac{\partial}{\partial P_y}(P_yf_0(P))dP_y=
P_yf_0(P)\Bigg|_{P_y=-\infty}^{P_y=+\infty}=0.
$$

Now in the second integral from (2.4) we will calculate internal integral
on $P_x $
$$
\int\limits_{-\infty}^{\infty}\dfrac{\partial}{\partial P_x}
\Big(\dfrac{f_0(P)}{\Omega-qP_x}\Big)
\dfrac{P_xdP_x}{\Omega-qP_x}=
$$
$$
=\dfrac{f_0(P)P_x}{(\Omega-qP_x))^2}\Bigg|_{P_x=-\infty}^{P_x=+\infty}-
\int\limits_{-\infty}^{\infty}\dfrac{f_0(P)}{\Omega-qP_x}
d\Big(\dfrac{P_x}{\Omega-qP_x}\Big)=
$$
$$
=-\Omega \int\limits_{-\infty}^{\infty}
\dfrac{f_0(P)dP_x}{(\Omega-qP_x)^3}.
$$

Thus, equality (2.5) becomes simpler
$$
j_x=-\dfrac{e^3E_y^2}{k_T^2m^2}q\int \dfrac{f_0(P)P_y^2d^3P}
{(\Omega-qP_x)^3}.
\eqno{(2.5)}
$$

Double internal integral from (2.5) in  plane $(P_y, P_z)$
it is calculated in polar coordinates
($P_y=\rho\cos \varphi,P_z=\rho\sin\varphi$)
$$
\int\limits_{-\infty}^{\infty}\int\limits_{-\infty}^{\infty}
e^{-P_y^2-P_z^2}P_y^2dP_ydP_z=\int\limits_{0}^{2\pi}\int\limits_{0}^{\infty}
\cos^2\varphi e^{-\rho^2}\rho^3d\varphi d\rho=\dfrac{\pi}{2}.
$$

Hence, the density of  longitudinal current is equal to
$$
j_x=\dfrac{e^3E_y^2Nq}{2k_T^2m^2v_T^3\sqrt{\pi}}
\int\limits_{-\infty}^{\infty}
\dfrac{e^{-P_x^2}dP_x}{(qP_x-\Omega)^3}.
\eqno{(2.6)}
$$

Выражение (2.6) представим в виде:
$$
j_x^{\rm long}=J_{c}(\Omega,q)\sigma_{l,tr}kE_y^2.
\eqno{(2.7)}
$$

В (2.7) $\sigma_{l,tr}$ -- продольно--поперечная проводимость,
$J_c(\Omega,q)$ -- безразмерная часть продольного тока,
$$
\sigma_{l,tr}=\dfrac{e}{p_Tk_T}\Omega_p^2=\dfrac{e\hbar}{p_T^2}
\Big(\dfrac{\hbar \omega_p}{mv_T^2}\Big)^2=\dfrac{e\hbar}{p_T^2}
\Big(\dfrac{\omega_p}{k_Tv_T}\Big)^2,
$$
$$
J_c(\Omega,q)=\dfrac{1}{8\pi\sqrt{\pi}}
\int\limits_{-\infty}^{\infty}
\dfrac{e^{-P_x^2}dP_x}{(qP_x-\Omega)^3},
$$
$\omega_p$ -- плазменная (лэнгмюровская) частота, $\Omega_p$ --
безразмерная плазменная частота,
$$
\omega_p=\sqrt{\dfrac{4\pi e^2N}{m}}, \qquad
\Omega_p=\dfrac{\omega_p}{k_Tv_T}.
$$

Let us present the formula (2.6) in an invariant form.
For this purpose we will introduce transversal electric field
$$
\mathbf{E}_{\rm tr}=\mathbf{E}-\dfrac{\mathbf{k(Ek)}}{k^2}=
\mathbf{E}-\dfrac{\mathbf{q(Eq)}}{q^2}.
$$

Now equality (2.7) we will present in coordinate-free form
$$
\mathbf{j}^{\rm long}= J_{c}(\Omega,q)\sigma_{l,tr}{\bf k}
\mathbf{E}_{\rm tr}^2=J_{c}(\Omega,q)\sigma_{l,tr}
\dfrac{\omega}{c}[{\bf E,H}].
$$

The integral (2.6) is calculated with use of known rule of Landau
as Cauchy type integral
$$
\int\limits_{-\infty}^{\infty}\dfrac{e^{-\tau^2}d\tau}{(q\tau-\Omega)^3}=
\lim\limits_{\varepsilon\to 0}
\int\limits_{-\infty}^{\infty}
\dfrac{e^{-\tau^2}d\tau}{(q\tau-\Omega+i\varepsilon)^3}=
$$
$$
=-i\pi \dfrac{1}{2q^3}\Big(e^{-\tau^2}\Big)''\Bigg|_{\tau=\Omega/q}+
{\rm V.p.}\int\limits_{-\infty}^{\infty}
\dfrac{e^{-\tau^2}d\tau}{(q\tau-\Omega)^3}.
$$\medskip

On fig. 1-3 we will present graphics of behaviour of the real
part of dimensionless quantity of density of electric current
$$
\Re J_c(\Omega,q)={\rm V.p.}\dfrac{1}{8\pi\sqrt{\pi}}
\int\limits_{-\infty}^{\infty}
\dfrac{e^{-\tau^2}d\tau}{(q\tau-\Omega)^3}.
$$
On fig. 4-6 we will present graphics of behaviour of the
imaginary part of dimensionless quantity of density of electric
current,

$$
\Im J_c(\Omega,q)=-\dfrac{1}{16 \sqrt{\pi}q^3}(e^{-\tau^2})''
\Bigg|_{\tau=\Omega/q}=\dfrac{1}{8\sqrt{\pi}q^3}e^{-(\Omega/q)^2}
\Big[1-2\Big(\dfrac{\Omega}{q}\Big)^2\Big].
$$

In case of small values of wave number from (2.6) it is received
$$
j_x=-\dfrac{e^3E_y^2(x,t)N}{2m^2\omega^3}\cdot k=
-e\Big(\dfrac{\omega_p}{\omega}\Big)^2\dfrac{kE_y^2}{8\pi m\omega}=.
$$
$$
=-e\Big(\dfrac{\Omega_p}{\Omega}\Big)^2\dfrac{kE_y^2}{8\pi m\omega}=-
\dfrac{e}{8\pi}\Big(\dfrac{\Omega_p}{\Omega}\Big)^2
\dfrac{kE_y^2}{\Omega p_Tk_T}
=-\sigma_{l,tr}\dfrac{kE_y^2}{8\pi\Omega^3}.
\eqno{(2.8)}
$$

\begin{center}
  \bf 3. Quantum Maxwellian plasma
\end{center}

Quantum Wigner distribution function was constructed in our work
\cite{Lat1} in general case
$$
f=f_0(P)+\dfrac{ev_T}{c\hbar}\mathbf{PA}\dfrac{f_0(\mathbf{P}+
\mathbf{q}/{2})-f_0(\mathbf{P}-\mathbf{q}/{2})}
{\omega-v_T\mathbf{kP}}+
\dfrac{e^2v_T^2(\mathbf{PA})^2}{2c^2\hbar^2(\omega-v_T\mathbf{kP})}\times
$$
$$
\Big[\dfrac{f_0(\mathbf{P+q})-f_0(P)}{\omega-v_T\mathbf{k(P+q}/2)}+
\dfrac{f_0(\mathbf{P-q})-f_0(P)}{\omega-v_T\mathbf{k(P-q}/2)}\Big]-
\dfrac{e^2\mathbf{A}^2}{4mc^2\hbar}\dfrac{f_0(\mathbf{P+q})-
f_0(\mathbf{P-q})}{\omega-v_T\mathbf{kP}}.
$$

By definition, the electric current density is equal
$$
\mathbf{j}(\mathbf{r},t)=e\int \mathbf{v}(\mathbf{r},\mathbf{p},t)
f(\mathbf{r},\mathbf{p},t)d^3v.
\eqno{(3.1)}
$$

Substituting in equality (4.1) obvious expression for speed
$$
\mathbf{v}(\mathbf{r},\mathbf{P},t)=
v_T\mathbf{P}-\dfrac{e \mathbf{A}(\mathbf{r},t)}{mc},
$$
and distribution function.

Leaving linear and square-law expressions concerning
vector potential of electromagnetic field, we receive
$$
\mathbf{j}=ev_T^3\int \Big[v_T\mathbf{P}f_1-
\dfrac{e}{mc}\mathbf{A}f_0(P)\Big]d^3P+
$$
$$
+ev_T^3\int \Big[v_T\mathbf{P}f_2-
\dfrac{e}{mc}\mathbf{A}f_1\Big]d^3P.
\eqno{(4.2)}
$$

Let us show, that the formula (4.2) for electric current density
contains two non-zero components: $ \mathbf {j} = (j_x, j_y, 0) $.
One component $j_y $ is linear on potential of an electromagnetic field and
are directed lengthways field. It is the known formula for
electric current density, so-called "transversal current".
The second  component $j_x $ is quadratical
on potential of  field also it is directed along the wave
vector. It is "longitudinal current".

The first composed in (4.2) is linear on vector potential
expression, and second is square-law. We will write out these composed in
obvious kind
$$
\mathbf{j}^{\rm linear}=ev_T^3\int \Bigg[
\dfrac{ev_T^2}{c\hbar}\mathbf{P(PA)}\dfrac{f_0(\mathbf{P+q}/2)-
f_0(\mathbf{P-q}/2)}{\omega-v_T\mathbf{kP}}-\dfrac{e\mathbf{A}}{mc}f_0(P)
\Bigg]d^3P
\eqno{(4.3)}
$$
and
$$
\mathbf{j}^{\rm quadr}=ev_T^3\int \Bigg[
-\dfrac{e^2v_T\mathbf{A(PA)}}{mc^2\hbar}
\big[f_0(\mathbf{P}+\dfrac{{\bf q}}{2})-
f_0(\mathbf{P}-\dfrac{{\bf q}}{2})\big]+
$$
$$
+\dfrac{e^2v_T^3\mathbf{P(PA)}^2}{2c^2\hbar^2}
\Big[\dfrac{f_0(\mathbf{P+q})-f_0(P)}{\omega-v_T\mathbf{k(P+q}/2)}-
\dfrac{f_0(P)-f_0(\mathbf{P-q})}{\omega-v_T\mathbf{k(P-q}/2)}\Big]-
$$
$$
-\dfrac{e^2v_T\mathbf{PA}^2}{4mc^2\hbar}\big[f_0(\mathbf{P+q})-
f_0(\mathbf{P-q})\big]\Bigg]
\dfrac{d^3P}{\omega-v_T\mathbf{kP}}.
\eqno{(4.4)}
$$

Expression (4.3) is linear expression of
the electric current, found, in particular, in our previous
work \cite{Lat1}. This vector expression contains only one
the component, directed along the electromagnetic
fields. Really, if wave vector to direct along an axis
$x $ i.e. to take $ \mathbf {k} =k (1,0,0) $, and potential electromagnetic
fields to direct along an axis $y $, i.e. to take
$ \mathbf {A} (\mathbf {r}, t) = (0, A_y (x, t), 0) $, from the formula (4.3)
we receive
$$
j_y^{\rm linear}=-\dfrac{e^2v_T^3A_y}{mcq}
\int\Big(\dfrac{f_0(P_x+q/2)-f_0(P_x-q/2)}
{P_x-\Omega/q}P_y^2+qf_0(P)\Big)d^3P.
\eqno{(4.5)}
$$

Here
$$
f_0(P_x\pm q/2)=\dfrac{N}{\pi^{3/2}v_T^3}e^{-(P_x\pm q)^2-P_y^2-P_z^2},\qquad
\Omega=\dfrac{\omega}{k_Tv_T}.
$$

Let us consider expression for an electric current (4.4),
proportional to  square of potential of  electromagnetic field.
Let us notice, that the first composed in this expression is equal to zero.
Hence, this expression becomes simpler
$$
\mathbf{j}^{\rm quadr}=ev_T^3\int \Bigg[
\dfrac{e^2v_T^3\mathbf{P(PA)}^2}{2c^2\hbar^2(\omega-v_T\mathbf{kP})}
\Big[\dfrac{f_0(\mathbf{P+q})-f_0(P)}{\omega-v_T\mathbf{k(P+q}/2)}+$$$$+
\dfrac{f_0(\mathbf{P-q})-f_0(P)}{\omega-v_T\mathbf{k(P-q}/2)}\Big]-
\dfrac{e^2v_T\mathbf{PA}^2}{4mc^2\hbar}\dfrac{f_0(\mathbf{P+q})-
f_0(\mathbf{P-q})}{\omega-v_T\mathbf{kP}}\Bigg]d^3P.
\eqno{(4.6)}
$$

Let us notice, that vector expression (4.6) contains one non-zero
the electric current component, directed along the wave
vector
$$
{j_x}^{\rm quadr}=
\dfrac{e^3v_T^2A_y^2}{2c^2m^2}\int \Bigg[
\Big[\dfrac{f_0(P_x+q)-f_0(P)}{qP_x+q^2/2-\Omega}+
\dfrac{f_0(P_x-q)-f_0(P)}{qP_x-q^2/2-\Omega}\Big]P_y^2
+
$$
$$
+\dfrac{f_0(P_x+q)-f_0(P_x-q)}{2}\Bigg]\dfrac{P_xd^3P}{qP_x-\Omega}.
\eqno{(4.7)}
$$

Let us lead to the form convenient for calculations, the formula (4.7) for
density of a longitudinal current.

Let us consider the first integral from (4.7). We will calculate the internal
integrals in a plane $ (P_y, P_z) $, passing to polar coordinates
$$
\int f_0(P_x\pm q)P_y^2dP_ydP_z=\dfrac{Ne^{-(P_x\pm q)^2}}{2v_T^3\sqrt{\pi}},
$$
$$
\int f_0(P_x\pm q)dP_ydP_z=\dfrac{Ne^{-(P_x\pm q)^2}}{v_T^3\sqrt{\pi}}.
$$

Thus, size of  generated longitudinal current into quantum plasma is equal
$$
j_{x}^{\rm quant}=
\dfrac{e^3NA_y^2}{4m^2c^2v_T\sqrt{\pi}}
\int\limits_{-\infty}^{\infty}\Bigg[
\dfrac{e^{-(P_x+q)^2}-e^{-P_x^2}}{qP_x+q^2/2-\Omega}+
\dfrac{e^{-(P_x-q)^2}-e^{-P_x^2}}{qP_x-q^2/2-\Omega}+
$$
$$
+e^{-(P_x+q)^2}-e^{-(P_x-q)^2}\Bigg]\dfrac{P_x dP_x}{qP_x-\Omega}.
\eqno{(4.8)}
$$

Let us transform the expression standing in integral (4.8). At first
let us pass from potential to intensity of the field
$A_y = - (ic/\omega) E_y $. We will receive
$$
\dfrac{e^3NA_y^2}{4m^2c^2v_T\sqrt{\pi}}=
-\dfrac{e^3NE_y^2}{4m^2v_T\omega^2 \sqrt{\pi}}=
-\dfrac{e\omega_p^2 E_y^2}{16\pi \sqrt{\pi}p_T\omega^2}=
$$
$$
=-\dfrac{e\Omega_p^2}{16\pi\sqrt{\pi} p_Tk_T}\dfrac{k}{\Omega^2q}E_y^2=
-\dfrac{\sigma_{l,tr}}{16\pi \sqrt{\pi}}\dfrac{k}{\Omega^2q}E_y^2,
$$
where the quantity of longitudinal--transversal conductivity
$\sigma_{l,tr}$ was introduced earlier: $\sigma_{l,tr}=e\Omega_p^2/(p_Tk_T)$.

Now equality (4.8) we will present in the form
$$
j_x^{\rm quant}=J_{\rm q}(\Omega,q)\sigma_{l,tr}kE_y^2=
J_{\rm q}(\Omega,q)\sigma_{l,tr}\dfrac{\omega}{c}E_yH_z.
\eqno{(4.9)}
$$

In (4.9) $J(\Omega,q)$ is the density of dimensionless
longitudinal current,
$$
J_{\rm q}(\Omega,q)=-\dfrac{1}{16\pi\sqrt{\pi} \Omega}
\int\limits_{-\infty}^{\infty}
\dfrac{[3q^2+2(q\tau-\Omega)^2-q^4/2]e^{-\tau^2}d\tau}
{[(q\tau-\Omega)^2-q^4/4][(q\tau-\Omega)^2-q^4]}.
\eqno{(4.10)}
$$

At calculation singular integral from (4.10), which not writing out
let us designate through $I (\Omega, q) $, it is necessary to take advantage
known Landau rule. Then
$$
I(\Omega,q)=\Re I(\Omega,q)+i\Im I(\Omega,q).
$$

Here
$$
\Re I(\Omega,q)={\rm V.p.}
\int\limits_{-\infty}^{\infty}
\dfrac{[3q^2+2(q\tau-\Omega)^2-q^4/2]e^{-\tau^2}d\tau}
{[(q\tau-\Omega)^2-q^4/4][(q\tau-\Omega)^2-q^4]},
$$
symbol $ {\rm V.p.} $ means a principal value of integral,
$$
\Im I(\Omega,q)=-\pi \dfrac{\Omega}{q^2}\Bigg\{-\dfrac{4}{q^2}
\Big[e^{-(\Omega/q+q/2)^2}-e^{-(\Omega/q-q/2)^2}\Big]+
$$
$$
+\Big(\dfrac{2}{q^2}+1\Big)\Big[e^{-(\Omega/q+q)^2}-e^{-(\Omega/q-q)^2}\Big]
\Bigg\}.
$$

Equality (4.10) for density of longitudinal current we will present in
the vector form
$$
{\bf j}^{\rm quant}=J_{\rm q}(\Omega,q)\sigma_{l,tr}{\bf kE}_{tr}^2,
$$
or
$$
{\bf j}^{\rm quant}=J_{\rm q}(\Omega,q)\sigma_{l,tr}\dfrac{\omega}{c}
[{\bf E,H}].
$$

Let us show, that at small values of wave number density
longitudinal current both in quantum and in classical plasma
coincide.

According to (4.7) at small $q $ after
linearization  $f_0 (P_x\pm q) $ and transition from vector potential to
electromagnetic field we receive
$$
j_x^{\rm quant}=-\dfrac{e^3Т N qE_y^2}{m^2v_T\omega^2\Omega\pi^{3/2}}
\int e^{-P_x^2}P_x^2d^3P=-\dfrac{e^3NqE_y^2}{2m^2v_T \Omega\omega^2}=
$$
$$
=-\dfrac{e\omega_p^2kE_y^2}{8\pi p_Tk_T \Omega\omega^2}=
-\dfrac{ekE_y^2}{8\pi p_Tk_T\Omega}\Big(\dfrac{\Omega_p}{\Omega}\Big)^2=
-\sigma_{l,tr}\dfrac{kE_y^2}{8\pi \Omega^3},
$$
that in accuracy coincides with expression (2.8) for the classical
plasma. These expressions we will copy in the vector form
$$
{\bf j}^{\rm long}=-
\sigma_{l,tr}\dfrac{1}{8\pi\Omega^3}\cdot{\bf k}{\bf E}_{tr}^2
=-
\sigma_{l,tr}\dfrac{1}{8\pi\Omega^3}\dfrac{\omega}{c}[{\bf E,H}].
$$

\begin{center}
\bf 3. Conclusions
\end{center}

In the present work the solution of Vlasov equation  for
collisionless Maxwellian plasmas is used. For the solution it the method
of consecutive approximations is used.

As small parametre the quantity of the vector
potential of electromagnetic field (or to it proportional
quantity of intensity of electric field) is considered.

At use of approximation of the second order it appears, that
the electromagnetic field generates an electric current directed
along the wave vector, and proportional to the size of square
of electric field.

\clearpage

\begin{figure}[t]\center
\includegraphics[width=16.0cm, height=10cm]{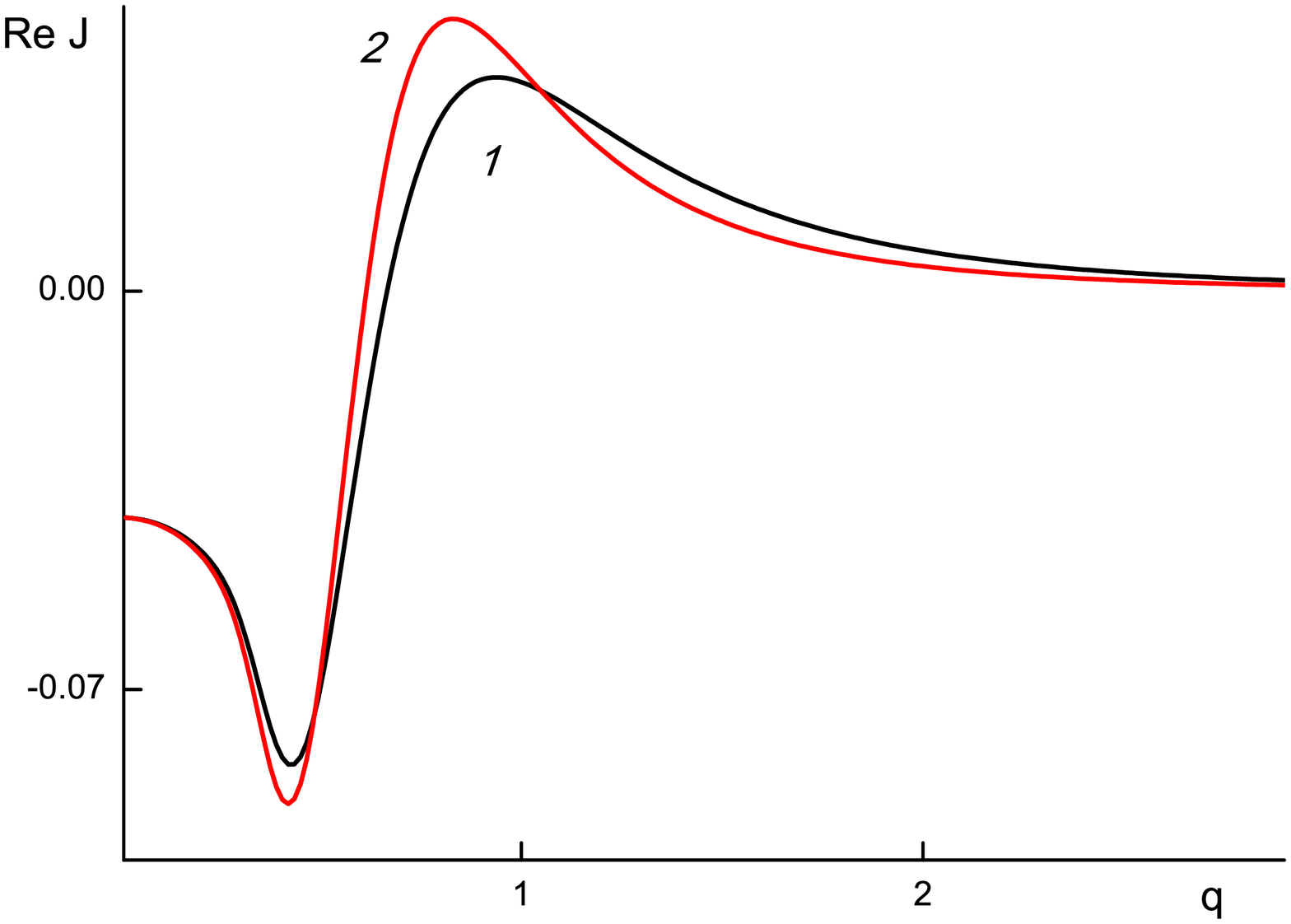}
\center{Fig. 1. Real part of longitudinal electric current density,
$\Omega=1$. Curves $1,2$ correspond to classical Maxwellian plasma and
classical Fermi---Dirac plasma, dimensionless
chemical potential $\alpha=0$.}
\includegraphics[width=17.0cm, height=10cm]{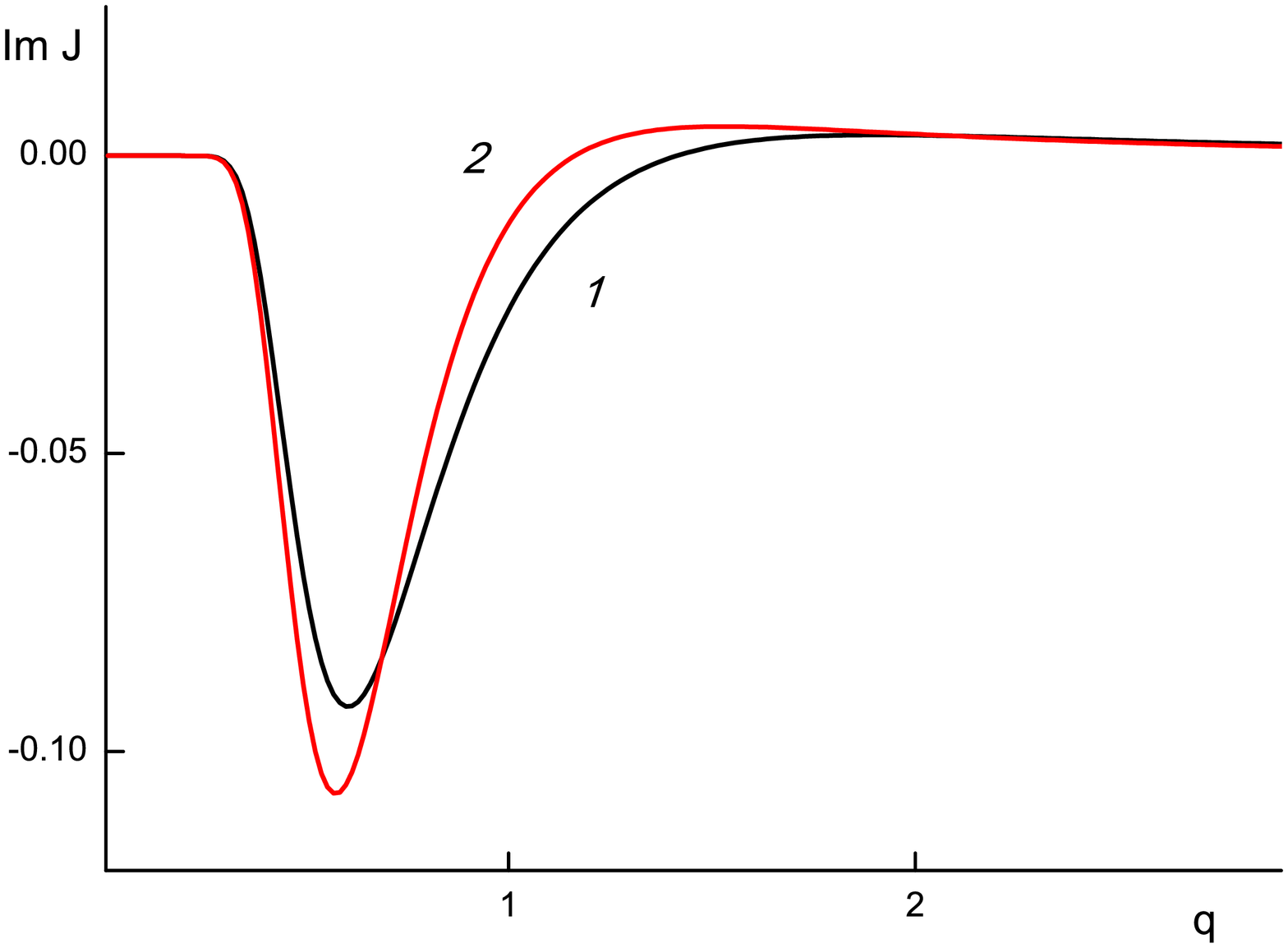}
\center{Fig. 2. Imaginary part of longitudinal electric current density,
$\Omega=1$.  Curves $1,2$ correspond to classical Maxwellian plasma and
classical Fermi---Dirac plasma, dimensionless
chemical potential $\alpha=0$.}
\end{figure}

\begin{figure}[t]\center
\includegraphics[width=16.0cm, height=10cm]{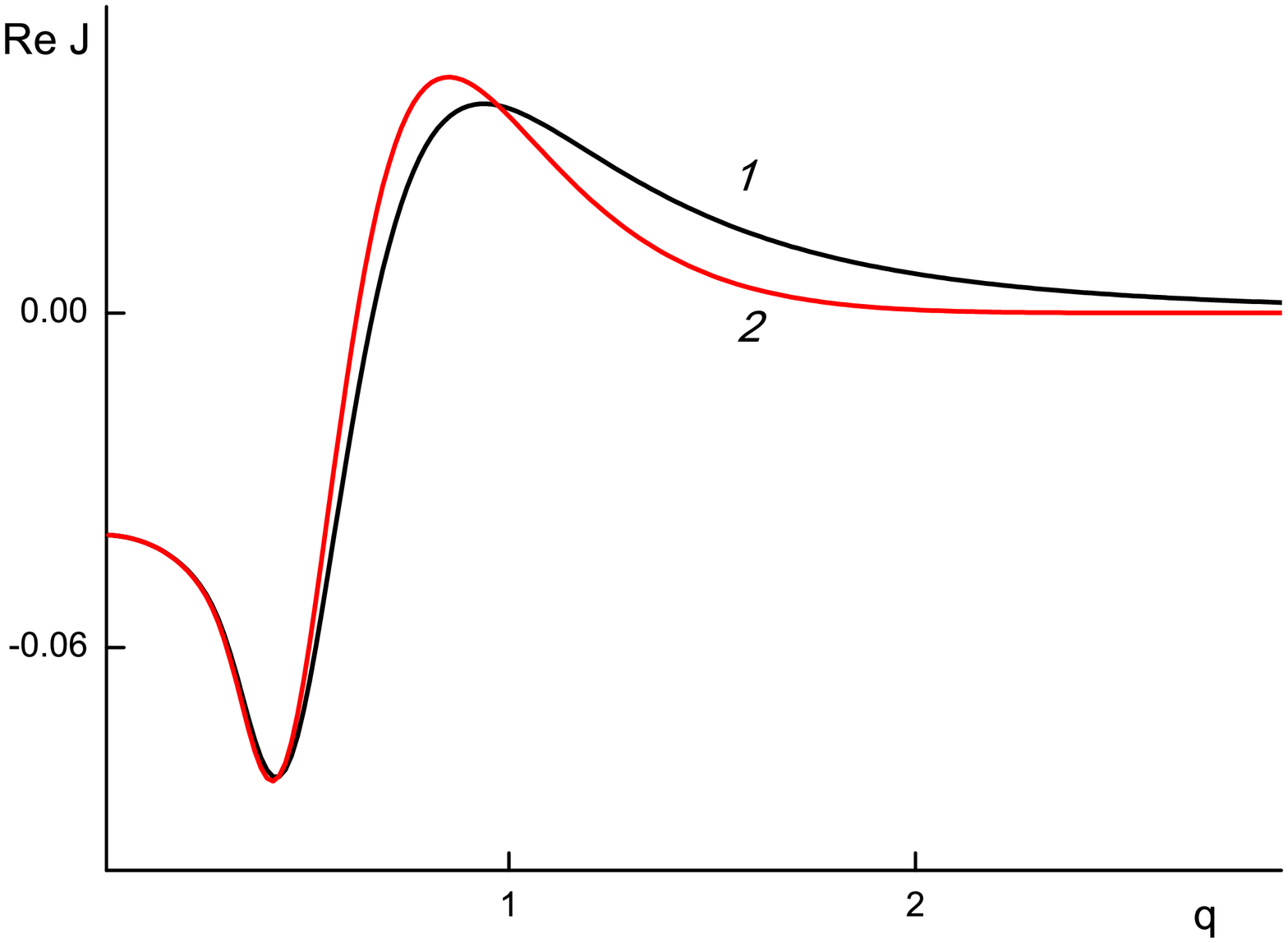}
\center{Fig. 3. Real part of longitudinal electric current density.
Curves $1,2$ correspond to Maxwellian classical and quantum plasma.}
\includegraphics[width=17.0cm, height=10cm]{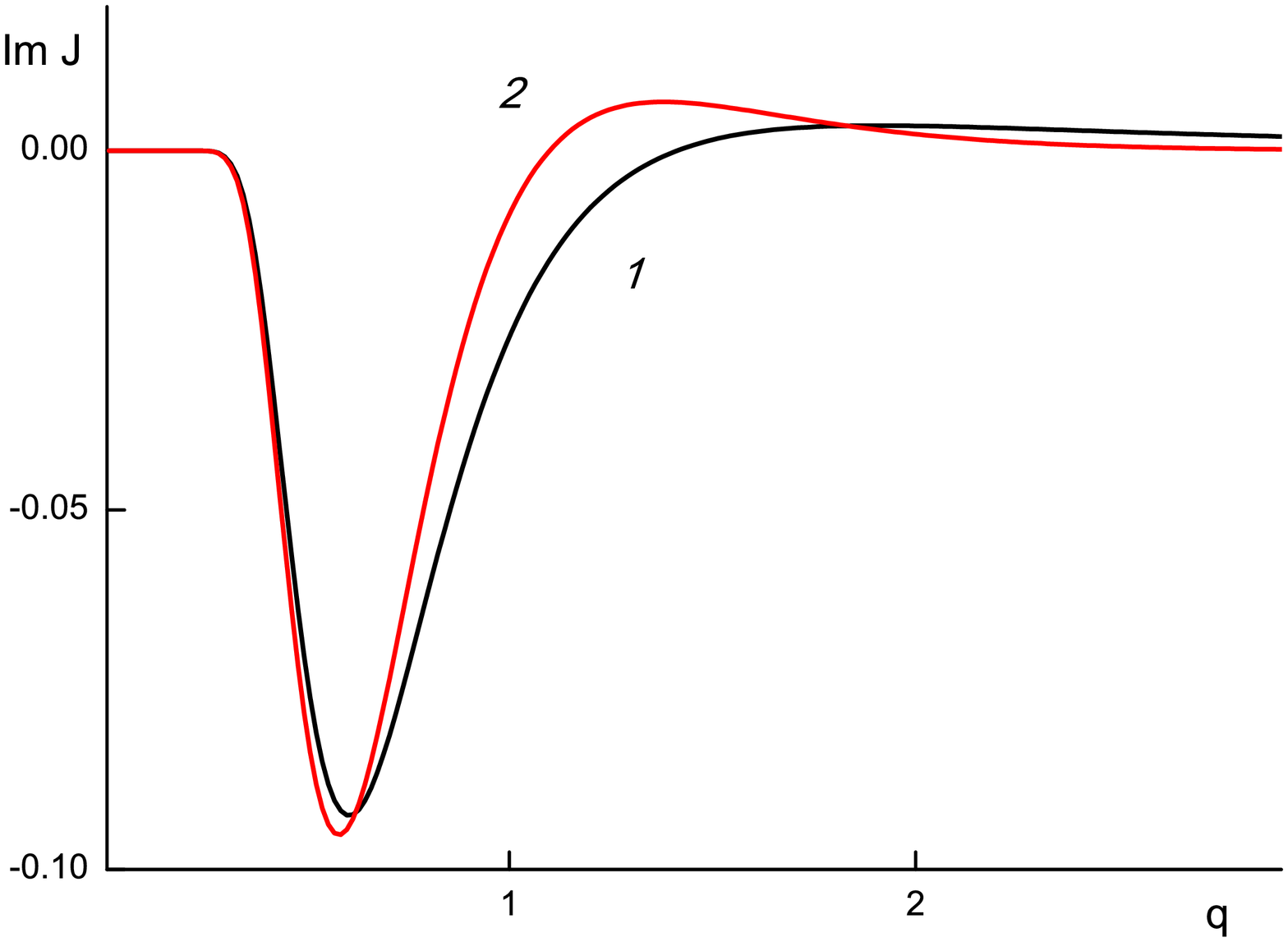}
\center{Fig. 4. Imaginary part of longitudinal electric current density.
Curves $1,2$ correspond to Maxwellian classical and quantum plasma.}
\end{figure}

\begin{figure}[t]\center
\includegraphics[width=16.0cm, height=10cm]{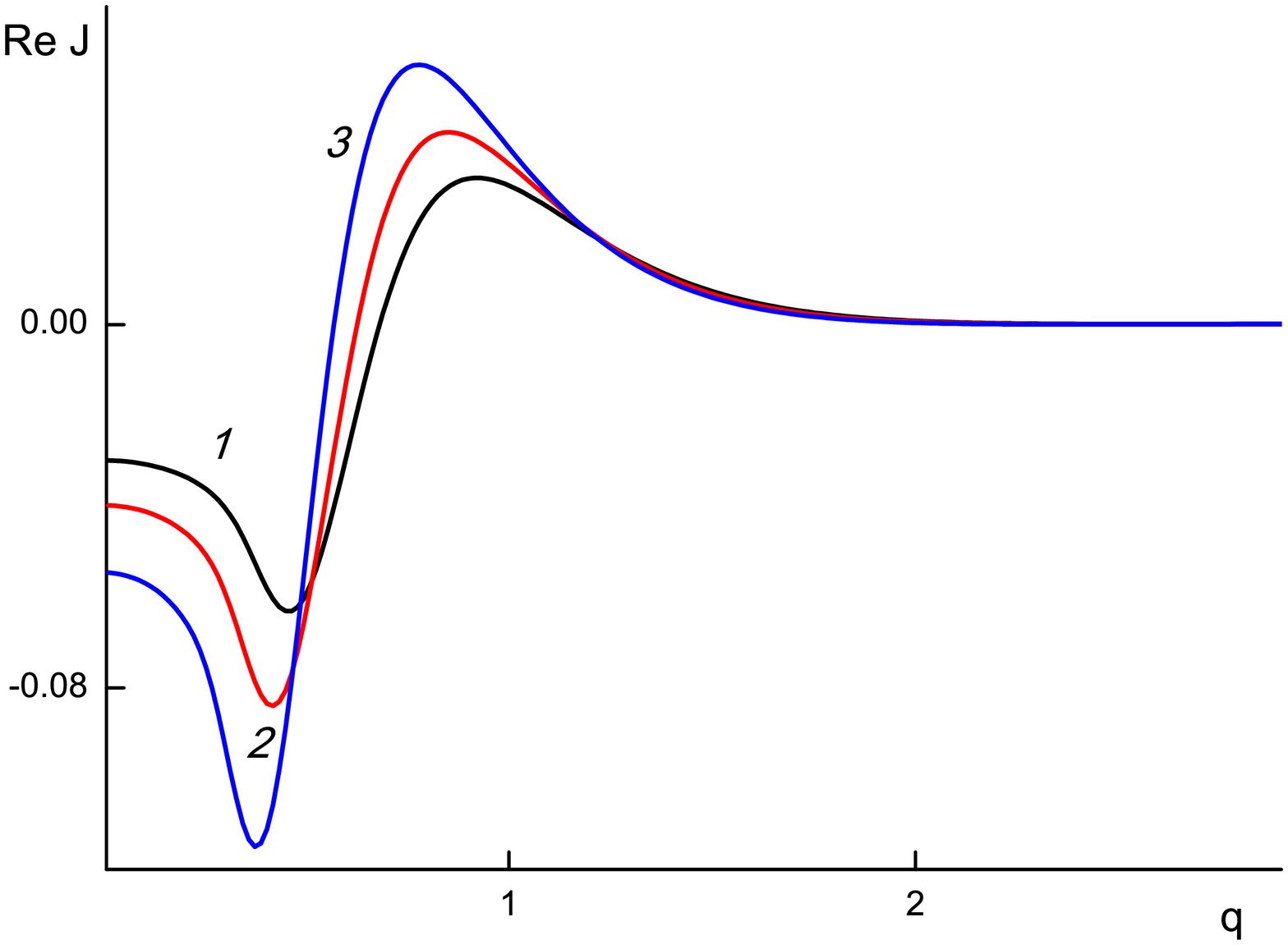}
\center{Fig. 5. Real part of longitudinal electric current density
of quantum Maxwellian plasma.
Curves $1,2,3$ correspond to values of dimensionless
frequency of electromagnetic field $\Omega=1.1,1,0.9$.}
\includegraphics[width=16.0cm, height=10cm]{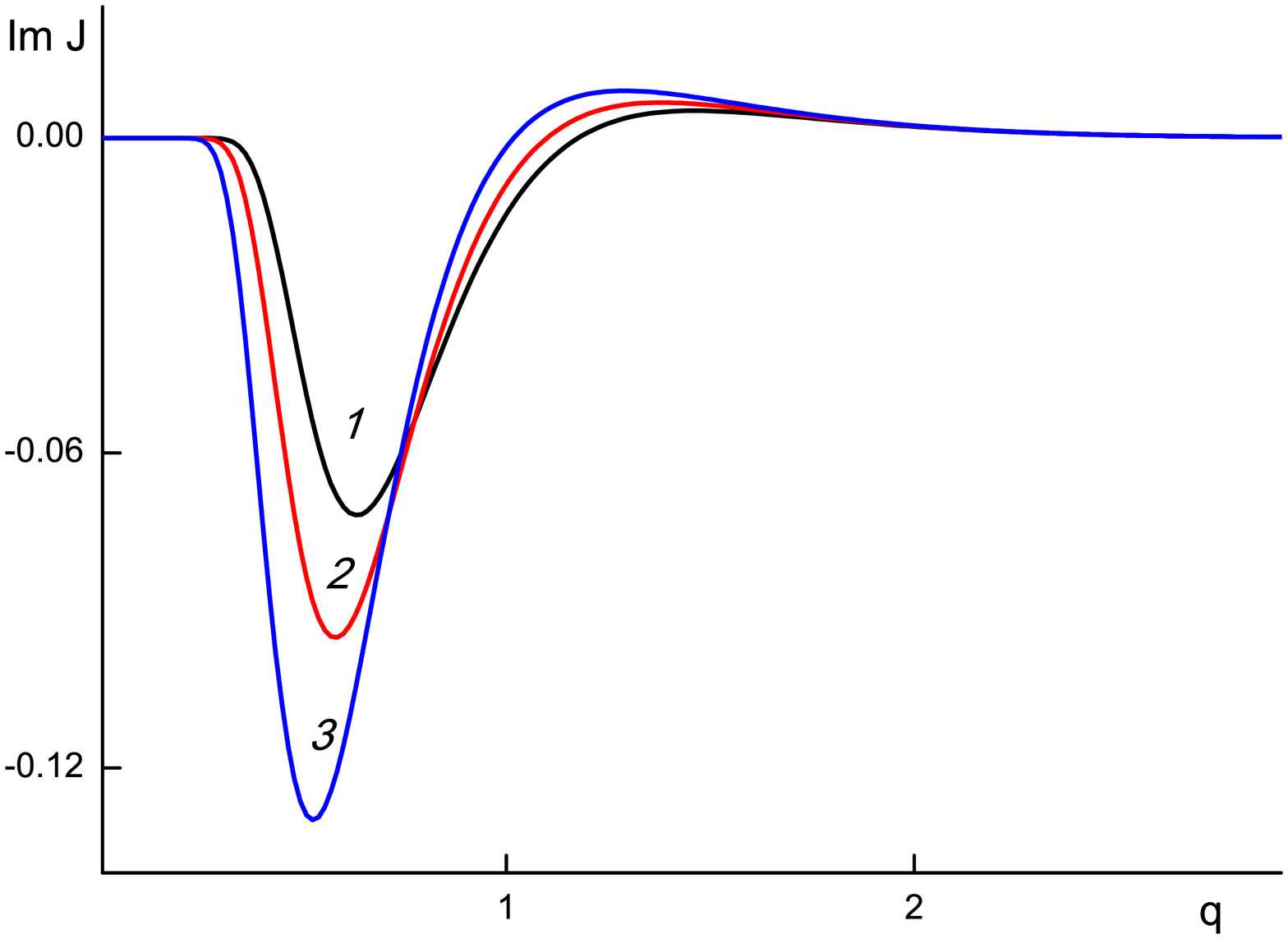}
\center{Fig. 6. Imaginary part of longitudinal electric current density,
of quantum Maxwellian plasma. Curves $1,2,3$
correspond to values of dimensionless
frequency of electromagnetic field $\Omega=1.1,1,0.9$.}
\end{figure}

\end{document}